\documentclass[twocolumn,showpacs,preprintnumbers,amsmath,amssymb,footinbib]{revtex4-1}

 \usepackage{CJK}
\usepackage{graphicx}
\usepackage{epstopdf,mathrsfs}
\usepackage{booktabs}
\usepackage{enumitem}

\begin{document}
\begin{CJK*}{}{}

\title{800 nm pumped SPDC from chirped crystals}

\author{X. S\'anchez-Lozano, C. Wiechers and J. L. Lucio}
\affiliation{Departamento de F\'isica-DCI, Universidad de Guanajuato, P.O. Box E-143, 37150, Le\'on, Gto., M\'exico.}
\date{\today}

\begin{abstract}
\noindent We consider crystal chirp effects on SPDC when pumping at 800 nm. The typical distribution produced in frequency-momentum space is a pop tab-like structure which turns out to be suitable for the implementation of versatile light sources. Our analyzes consider the effect of internal and external parameters in the process; in the former we include the crystal chirp and length, while in the latter temperature, as well as pump chirp and beam properties. We report evidence of the appropriateness of SPDC from chirped crystals to manipulate the frequency and transverse momentum properties of the light produced. We briefly comment on potential usefulness of the types of light produced, in particular for quantum information applications.
\end{abstract}

\maketitle

\end{CJK*}

\section{Introduction}

\noindent Spontaneous parametric down-conversion (SPDC), the process in which single photons from a pump beam are annihilated in a second-order ($\chi^{(2)}$) non-linear crystal, leading to the generation of signal and idler photon pairs, has shown remarkable flexibility  to produce photon pairs with a wide range of  spatial and spectral characteristics. Indeed, careful design of the attributes of the nonlinear crystal and the pump beam result in photon states with fitted properties to specific needs. The configuration of the experimental setups, the degrees of freedom under consideration and their interrelation are relevant for the produced state; and the ability to control it. Although a wide range of studies have been devoted to this topic, there is still room for extending the tools and procedures to improve the quality and  specificity of the two photon quantum state.\\
\vspace{0.15cm}

\noindent Already in the early stages of SPDC description, it was realized that the two photon state carries information about the angular spectrum of the pump beam \cite{PRA57Monken}, and since then detailed analyses of  transverse properties of the field generated in that process \cite{Cosme,yo2012,Ansari,LP23Bobo} have been reported. The pump-signal-idler momenta configuration and  the geometry of the non-linear media play a pivotal role. Thus, periodic po\-ling has been systematically used to obtain quasi-phase-matching (QPM) as a way to control the spatial mode function, leading to correlations between the output photons, opening the possibility of producing entanglement in an infinite dimensional Hilbert space  \cite{OP29-Torres, PRL101-Yu, PRA86-Svozilik}.\\

\noindent In order to set the context of our work, it is convenient to establish the characteristics of the generated light, since although  the two photon state produced in SPDC  have been widely used, their features have not been fully exploited. The second order non-linear interactions ena\-ble the three photon interaction which can be depicted as one body decaying into two particles. When plane wave pump in an unbounded free space is considered, the phase matching condition leads to a Dirac delta expressing momentum conservation. However due to the crystal finiteness, the Dirac delta converts into a {\it{sinc}} function (for finite homogeneous crystal and periodic poled crystal), or in a combination of error functions (aperiodic poled crystal). The model leads to a sixfold differential amplitude, the six components of the idler and signal photon momenta, in which one momentum component can be traded by the frequency. Thus, the state produced in SPDC  depends upon six variables. Typically, density plots as function of two variables are reported, in our case, two of the remaining variables are numerically integrated and particular values are fixed for the other two (equivalent to set a filters).  In this way, correlations between the energy of one of the photons and the momentum of the companion photon are expounded. This statement can be paraphrased in terms of the correlation between the spectrum (frequency) of one of the photons produced and the position (fixed by the transverse momentum) of the second photon. An alternative way to present the results is in terms of the corresponding Fourier transform, and then in place of the frequency dependence, the state depends on time and instead of the momentum, on the position.\\

\noindent An important feature of the process under consideration is the factorability, which refers to the possibility that the pumping properties and kinematical configuration of the actual implementation of SPDC is such that  the dependence on the signal, idler and pump photon momenta-frequency (position-time) can be factored so that the state depends on two particular combinations of these variables, but are independent of each other.In fact, a number of important results have been reported, among which we list the following since they are relevant for our discussion: \\

\noindent {\bf a)} The relations between frequency and spatial properties depend on the geometry of de SPDC configuration, including the characteristics of the pump and any fil\-te\-ring process  \cite{NJoP10-Osorio}.\\
\noindent {\bf b)} The coupling between frequency and transverse momentum entail correlations in these variables, so that it is wise to consider simultaneously spatial and temporal characteristic \cite{NJoP10-Osorio, PRL109-Jedrkiewicz}. \\
\noindent {\bf c)} The production of non-factorable X shape structure of spatio-temporal state has been demonstrated \cite{PRL102-Gatti, PRA82-Brambilla, PR85-Brambilla, PRL109-Jedrkiewicz, IJQI12-Gatti}, which opens the possibility of cus\-to\-mi\-zing the temporal properties by acting on the spatial degrees of freedom of the twin photons \cite{PRA81-Caspani}. This has been achieved using Beta Barium Borate (BBO) crystals and even more, it was shown that a transition from the production of non-factorable to factorable states can be attai\-ned modifying the SPDC configuration.\\
\noindent {\bf d)} Using periodically poled crystals is possible to attain high flux of photon-pairs, due to the strong non-linearity consequence of the QPM \cite{PRA73-Carrasco}.\\
\noindent {\bf e)} Including chirp in the crystal has an important effect on the spectrum \cite{PRA80-Svozilik, yo2015}, photon pairs are generated in a wider spectral bandwidth. This corresponds to a sharply localized state in time domain \cite{PRL98-Harris,PRL100-Nasr,yo2015}. It is also known that the larger the chirp the bigger photon-pair generation rate \cite{PRA2011-Perina}.\\
\noindent {\bf f)} There are two aspects whose relevance to the spectrum properties has been emphasized, one is the role played by the concave nature of the phase matching condition and the other is the relevance of the detection technique which should be able to resolve the spectrum produced \cite{PRA81-Caspani}.\\

\noindent In this work we consider SPDC from a chirped crystal with the aim of exploring the robustness of the phenomena observed as well as the versatility of SPDC as a light source. Some of the previously cited analyses include approximations, in the pumping properties or structure, in the phase matching condition and in the way crystal chirp is treated, among others. Here we avoid such approximations, and at the same time study the effect of external parameters on the two photon state properties. Thus, we consider pump chirp (can be implemented u\-sing an optical fiber) and temperature effects, besides the conventional parameters as the crystal length and chirp. Our results provides evidence of the appropriateness of SPDC from chirped crystals to manipulate the spectral and transverse momentum (time and spatial) properties of the two photon state.\\

\noindent It should be clear that photon pairs may be entangled in different photonic degrees of freedom, including  time, transverse position, frequency, momentum and po\-la\-ri\-za\-tion. Since special spectral and spatial properties of the entangled two photon states is required for efficient information encoding, then a detailed analyses of correlations is mandatory for security reasons and because such correlations open the possibility to ma\-ni\-pu\-late one degree of freedom by acting on a different one. Thus, for example, it is possible to manipulate temporal properties of the twin photons by acting on their spatial degrees of freedom. A broad kind of applications require light source, here it is enough to say that narrow and ultra-narrow band sources as well as broadband sources in the femtosecond range can be produced within the framework analyzed in this work.\\

\noindent The article is organized as follows. In Sec. II we set our conventions, state the assumptions used in the des\-crip\-tion of SPDC from a chirped non-linear crystal and write down the formulae used to perform the numerical calculations. In section III we report the comparison of the spectra obtained when different parameters of pump and/or non-linear crystal are used. The main cha\-rac\-teris\-tics and potential applications of different spectra are briefly described. In the final section we summarize our results and draw the conclusions.

 \section{ASSUMPTIONS AND FORMULATION}

\noindent Following the standard perturbative approach, where the state describing the SPDC process is approximated to $|\Psi\rangle = |0\rangle + \eta|\Psi_2\rangle + \dots$, where $\eta$ is related to efficiency; the two photon component $|\Psi_2\rangle$ is given by:
\begin{eqnarray}
|\Psi_2\rangle = \int d\vec{k}_i d\vec{k}_s  f(\vec{k}_s, \vec{k}_i)  |\vec{k}_s\rangle |\vec{k}_i\rangle,\label{psi2}
\end{eqnarray}
where $|\vec{k}_\mu\rangle=\hat{a}^\dag_\mu(\vec{k}_\mu)|0\rangle$ with $\mu~\epsilon ~\{s, i\}$ and $|0\rangle$ stands for the vacuum state. $f(\vec{k}_s, \vec{k}_i)$ is the joint amplitude function (JAF),  and is given by the product of two quantities with clear physical interpretation $f(\vec{k}_s, \vec{k}_i) =\Phi(\vec{k}_s, \vec{k}_i) \gamma(\omega_s, \omega_i)$. The pump envelope function (PEF) is described by $\gamma(\omega_s, \omega_i)$, whereas $\Phi(\vec{k}_s, \vec{k}_i)$ stands for the phase matching function (PMF). We consider non-collinear SPDC, where the pump, signal and idler photons propagate along different direction, the am\-pli\-tude of the ge\-ne\-ra\-ted entangled-photon pairs can be controlled through the modification of the pumping spectral shape \cite{NJoP10-Osorio}.
\vspace{0.15cm}

\noindent The characteristics of the medium where the non-linear interaction takes place are incorporated in the PMF:
\begin{eqnarray}
\Phi(\vec{k}_s, \vec{k}_i) = \int dV \textrm{d}(\vec{r})\tilde{\alpha}(\vec{r},\omega_s + \omega_i)  e^{-i(\vec{k}_s+ \vec{k}_i)\cdot \vec{r}}.\label{pm}
\end{eqnarray}
\noindent In this expression $\tilde{\alpha}(\vec{r},\omega_p)$ includes the spatial distribution of the pump and $\textrm{d}(\vec{r})$ portrays the crystal non\-li\-nea\-rity. Note that, in free space where d$(\vec{r})=1$ and for a plane wave $\tilde{\alpha}(\vec{r},\omega_p) \sim e^{i\vec{k_p} \cdot \vec{r}}$ the PMF reduces to a delta function that enforce momentum conservation. In the case of a non-linear crystal the function describing the non-linearity is approximated to $\textrm{d}(z) \propto e^{-i[K_0 + D(z_0 + z)]z}$. This expression emerges after considering the Fourier series of d($z$) and realizing that in virtue of the phase matching condition, the dominant contribution arises from the lowest (m=1) term \cite{yo2015}.

\noindent So far we have described the conventional formulation of the theoretical description of photon pair production through SPDC. Since analytical results and their conclusions can be reached only for restricted conditions, we prefer to rely on a numerical analysis of the two photon state pro\-per\-ties. So we start by listing the assumptions we make and the conditions we incorporate in our calculations:
\vspace{0.15cm}

\noindent \textbf{Non-linear crystal}. The use of a poled lithium niobate (LiNbO$_3$) crystal, described with a $z$-dependent second order susceptibility $\chi^{(2)}$, including periodic $z$ variation and crystal chirp. QPM is cha\-rac\-te\-ri\-zed by the spatial frequency  $K_0=2 \pi / \Lambda_c$, with $\Lambda_c$ the crystal period. When a li\-near chirp -quantified by $D$- is included in the crystal, the fixed grating spatial frequency $K_0$ is replaced by a function that grows linearly with the position
\begin{eqnarray}
K(z)=K_0+D(z_0+z),\label{K}
\end{eqnarray}
\noindent which is equivalent to introduce a \textit{local} period $\Lambda(z)= 2\pi/[K_0 + D(z_0 + z)]$ \cite{yo2015}. The reference coordinate $z_0$ is defined in the position where $\Lambda(z_0)=\Lambda_c$. For pumping at 800 nm and room temperature the central period is $\Lambda_c$ = 20.33 $\mu$m. We consider beam propagation along $z$ where a non-linear chirped crystal of length $L$ is placed between $z = A$ and $z = A+L$. Our numerical calculations are performed for distinct crystal lengths and temperatures (implying changes in the crystal central period $\Lambda_c$, such that if this is changed directly the results are similar \cite{JOSAB2013-Lerch}), or equivalently for distinct $z_0$ values, which may lie inside or outside the crystal.
\vspace{0.15cm}

\noindent \textbf{Pump properties}. Ultrafast pulsed pumping at the wavelength $\lambda_{pc} = 800$ nm with a Gaussian envelope func\-tion (PEF) with bandwidth $\sigma$, related directly to the full width at half maximum by FWHM $= 2\sqrt{2 \log 2}\sigma$; and Gaussian spatial distribution of waist $(W_x^2 + W_y^2)^{\frac{1}{2}}$; including also realistic time dependence of frequency (pump chirp $\beta$) \cite{yo2012}. In contrast with the monochromatic case, de\-pen\-ding upon the FWHM, the coherent addition of different pump spectral components plays a central role in the SPDC spectrum shape.

\noindent All three waves (pump, signal, and idler) are copolarized, in this way the largest elements of the non-linear susceptibility $\chi^{(2)}$; the diagonal, are involved, ensuring maximum efficiency. Thus we assume  type 0 phase-matching and that the polarization of all fields is extraordinary.
\vspace{0.15cm}

\noindent \textbf{Kinematics}. The quantum amplitude describing SPDC involves six variables. Under the assumption of a non-collinear geometry, our procedure amounts to integrate over two variables, fix two transverse momentum components, one for each photon, leaving the amplitude as a function of two momentum components (alternatively frequencies or its Fourier transform: position-time). Besides the phase matching condition, we do not place further constraints on the kinematics, \textit{i. e.} we include the complete correlation range for $k_\mu$ and $\omega_\mu$.

\begin{figure}[h!]
\begin{center}
\centering\includegraphics[height=3cm, width=8.2cm]{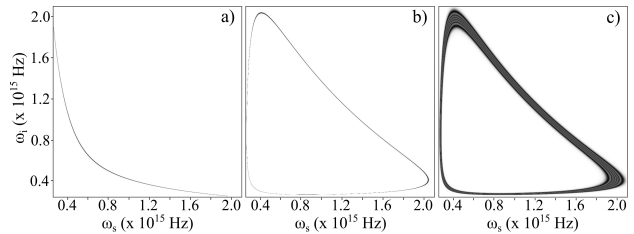}
\end{center}
\par
\caption{Phase matching function for different non-linear crystal with negative correlation in frequencies ($\beta=0$, $L$ = 5000 $\mu m$, FWHM = $0.01  ~\mu m$ and $W = 100 ~\mu m$): a) BBO - Type I, b) PPLN - Type 0 and c) Chirped PPLN - Type 0 ($D$ = $3 \times 10^{6} ~\mu m^{-2}$).} \label{pmf}
\end{figure}

\noindent Observables associated to SPDC are sensitive to the PMF since it strongly influence the shape of the two photon spectrum. Fig.\ref{pmf} shows the regions in the $\omega_s, \omega_i$ plane -shaded zones- compatible with the phase matching condition for three different non-linear media. The figures correspond to the case of a BBO crystal, a periodic crystal and a linearly chirped crystal. As compared to the sample BBO, periodicity in the crystal adds faint regions nearly parallel to the axes, while chirp effect broads the kinematical region permitted by the pe\-rio\-dic crystal, so that the larger the absolute value of the chirping pa\-ra\-me\-ter $D$, the broader the allowed ki\-ne\-ma\-ti\-cal zone.
\vspace{0.15cm}

\noindent Observables associated to SPDC are derived from the JAF. The following expression arises when the crystal boundaries are located at $A=-L$:
\begin{eqnarray}
\nonumber && f(\vec{k}_s, \vec{k}_i) = (-1)^{\frac{3}{4}}A_0g_{si}\pi^{\frac{3}{2}} W_xW_yL\sqrt{\frac{1}{4\xi}}e^{-\frac{\Delta  \omega^2}{\sigma^2}}e^{i\beta\Delta \omega^2} \\  &&\times e^{\frac{1}{4}[-\omega_{0}^2 + \frac{i\rho^2}{\xi}]}  (\mathrm{erf}[\frac{(-1)^\frac{1}{4}\rho}{2\sqrt{\xi}}] -\mathrm{erf}[\frac{(-1)^\frac{1}{4}(\rho - 2\xi)}{2\sqrt{\xi}}]),\label{jaf}
\end{eqnarray}
\noindent $A_0=2i/\pi$, $g_{si}$ is field amplitude, erf($z$) denotes the error function, the strength of the pump chirp is described by $\beta$ \cite{yo2012}, $\rho = -x_t + r\xi$, $W_\nu$ stands for the pump waist in direction $\nu~\epsilon~\{x, y\}$ and $k_t^2=k_x^2+k_y^2$, and
\begin{eqnarray}
\Delta\omega&=&\omega_s+\omega_i-\omega_{pc}  \qquad k_\nu=k_{\nu_s} + k_{\nu_i} \nonumber \\
\omega^2_0&=&\sum_{\nu=1}^2 k_\nu^2 W_\nu^2  \hspace{0.75cm} \qquad \xi=DL^2  \\
x_t &=& L\Delta k_t= L(k_p -k_{z_s}-k_{z_i} -2\pi/\Lambda_c - k_t^2/2k_p)  \nonumber
\nonumber
\end{eqnarray}
\noindent where we introduced the dimensionless quantity $r = z_0/L$, $x_t$ corresponds to an undimensional phase mismatch of a periodic crystal. Eq. \ref{jaf} reduces, in the appropriated limit, to expressions previously reported. Thus in the $\xi\rightarrow 0$ limit it reduces to the JAF for an unchirped QPM crystal, for  the collinear case \cite{yo2015} and with the ne\-ce\-ssa\-ry simplifications, we obtain Harris expressions \cite{PRL98-Harris}.
\vspace{0.15cm}

\noindent Results for the observables are attained from the joint probability distribution, the modulus squared of the JAF. Thus, Eq. \ref{jaf} allow us to obtain predictions re\-gar\-ding the two photon spectrum, or two variables asso\-cia\-ted to a single photon, but also the marginal distribution associated to a single variable. Our approach is fully numerical; and results are presented in plots, and we find convenient to report the joint probability distribution, the single photon spectrum and the Fourier transform to obtain temporal and spatial behavior.

\noindent To facilitate the presentation these quantities are denoted by  $J(v_1,v_2)$, $S(v)$ and $\mathcal{F}J(x_s, t_i)$ and are given by:
\begin{eqnarray}
\nonumber  \hspace{-0.7cm}J(k_{x_s}, \omega_i) &&=\int \hspace{-0.2cm}dk_{x_i} d\omega_s| f(\omega_s, \omega_i, k_{x_s}, k_{x_i}, k_{y_s}, -k_{y_s})|^2 \nonumber \\ S(\omega_i) &&=\int dk_{x_s} J(k_{x_s}, \omega_i), \\
\mathcal{F}J(x_s, t_i) &&=\frac{1}{(2\pi)^2} \int dk_{x_s} d\omega_i J(k_{x_s}, \omega_i) e^{-i(\omega_i t_i + k_{x_s} x_s)},
\nonumber  \label{jpd}
\end{eqnarray}
with $k_{y_s}=0$ and the joint amplitude normalized so that $\int d\vec{k}_s d\vec{k}_i |f(\vec{k}_s, \vec{k}_i)|^2=1$.
\vspace{0.15cm}

\noindent As to the degrees of freedom, we will focus on the fo\-llo\-wing variables: spectra $\omega_s, \omega_i$, transverse momenta $k_s, k_i$, position and time. As far sensitivity of the photon pair spectrum on experimentally controllable parameters, we considered different values for: crystal length $L$, the chirp parameter $D$, pump chirp $\beta$, FWHM and waist of Gau\-ssian beam and temperature.

\section{Spectra and its properties}

\noindent The probability that a given kinematical configuration in the SPDC actually materialize strongly depends on the PMF. Changes in the favored kinematical region depend upon the characteristics of the non-linear media and are schematically appreciated in Fig.\ref{pmf}, where a two dimensional ($\omega_s,\omega_i$) landscape is presented. Particularly relevant for our discussion is the effect produced by the crystal chirp, which besides broadening the allowed region generates bands  far from the vicinity where simple non-linear crystals (without chirp) have access.

\begin{figure}[h!]
\begin{center}
\centering\includegraphics[height=2.3cm, width=8cm]{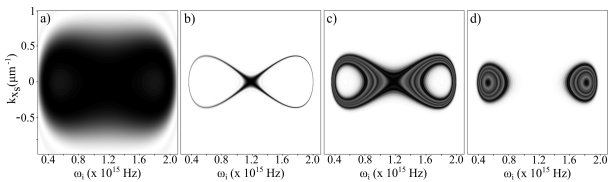}
\end{center}
\par
\caption{$J(k_{x_s}, \omega_i)$ for different non-linear crystals ($\beta=0$, $W_{x} = W_y = 100 ~\mu m$): a) $L$= 50 $\mu$m, b) FWHM = $0.001 ~\mu$m and $L$ = 5000 $\mu$m ($D=0$), c) FWHM = $0.001 ~\mu$m,  $L$ = 5000 $\mu$m and $D$= $2 \times 10^{-6} ~\mu m^{-2}$ and d) $T$ = 230$^\circ$C,  $L$ = 5000 $\mu$m, FWHM = $0.001 ~\mu$m for $D$= $2 \times 10^{-6} ~\mu m^{-2}$.} \label{varias}
\end{figure}

\noindent In Fig.\ref{varias} we collected spectra representative of our results and that can be described as the pop-tab like structure. The characteristics of this structure, such as the central symmetric shape and the ring like aspect in the extremes of the frequency distribution, are maintained with changes in the relative dimensions (ring radius and width) and intensity (see Fig.\ref{varias}c, d). The bottom line is that the use of near-infrared  wavelengths, chirp in the crystal and its combination with parameters externally controlled opens the way to a variety of possibilities once the  full phase space is considered.

\noindent SPDC will result in different spectra depending upon the experimental configuration. Before entering details regarding the two photon $-J(k_{x_s}, \omega_i)-$ and single photon spectrum $-S(\omega_i)-$, we start describing the results obtained. Fig.\ref{varias}a shows a spectrum whose characteristic is a broadband of uncorrelated pair of photons. No matter the frequency of the idler photon produced, there will be associated signal photons covering a whole band of momenta, while the marginal, both in frequency and momentum, display a flat supercontinuum spectrum.  \\

\noindent The pumping wavelength and the FWHM affect the spectrum. For the pumping considered in this work the spectrum shape produced with a non-linear crystal without chirp is depicted in Fig.\ref{varias}b. Salient features of the pop-tab structure are the correlation among the spectral profile and the transverse momentum; and the versatility it entails. Adaptability in the sense that it is possible to get a single or two spectral or transverse momentum peaks (frequency or transverse momentum selection); also a wide spectrum or an efficient source at a given wavelength. The limitation of this approach are set, among others, by kinematics. Fig.\ref{varias}c shows that the allowed kinematical regions can be sharpened, however the a\-nu\-lar structure keeps appearing. Thus for example, X shaped spectrum can be obtained through SPDC only for some combination of chirp and pump wavelength values.

\noindent  Fig.\ref{varias}d depicts a disconnected spectrum. In this case the central  symmetric shape of the spectrum is not part of the allowed kinematical region. The spectra include two frequency regions separated by domains where SPDC is not achievable or is strongly suppressed. In each of these two regions the pair of photons produced are uncorrelated. The marginal of this spectrum shows that for a given signal transverse momentum, there are two idler frequency peaks.

\begin{figure}[h!]
\begin{center}
\centering\includegraphics[height=4.5cm, width=8.0cm]{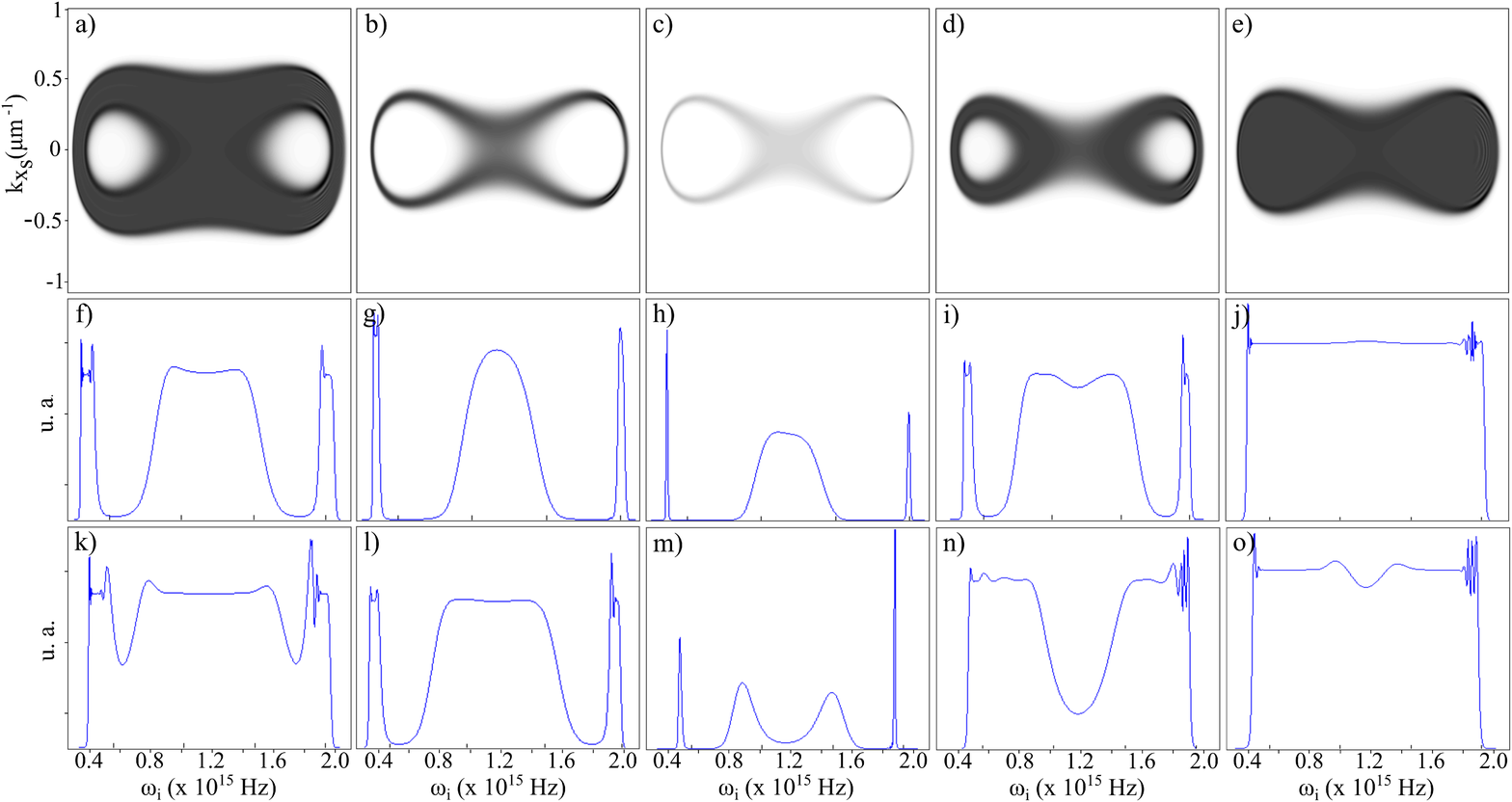}
\end{center}
\par
\caption{First row, $J(k_{x_s}, \omega_i)$ for different chirp in non-linear crystal ($\beta=0$, L = 5000 $\mu m$, FWHM = $0.01  ~\mu m$ and $W_{x} = W_y = 100 ~\mu m$): a) $D=-5 \times 10^{-6} ~\mu m^{-2}$, b) $-2 \times 10^{-6} ~\mu m^{-2}$, c) $0$, d) $2 \times 10^{-6} ~\mu m^{-2}$ and e) $5 \times 10^{-6} ~\mu m^{-2}$. Second and third row, the single photon spectrum for $k_{x_s}=0$ and $k_{x_s}=0.25 ~\mu m^{-1}$, respectively.} \label{chirp}
\end{figure}

\noindent In this work then, we analyze the effect of di\-ffe\-rent pa\-rame\-ters on the two photon spectrum. We consider degenerate non-collinear down conversion from a poled LiNbO$_3$, for three different lengths $L_1=50, L_2=500$ and $L_3= 5000~ \mu m$, with $z_0$ = 0.5 $L$ and distinct va\-lues of the crystal chirp $D=-5 \times 10^{-6} ~\mu m^{-2}, -2 \times 10^{-6} ~\mu m^{-2},  0, 2 \times 10^{-6} ~\mu m^{-2}$ and  $5 \times 10^{-6} ~\mu m^{-2}$. As far as the pum\-ping is concerned, we assume Gaussian pumping at 800 nm while for the FWHM we consider the values 0.001, 0.01 and 0.1   $\mu m$. The pump beam waist is fixed to $W_{x} = W_y = 100 \mu m$, since we found that the ($k_{x_s}, k_{x_i}$) correlation is sensitive and is enhanced with changes in the waist, but the  frequency-momentum correlations show very weak sensitivity to this parameter. As to the external parameters, we evaluated the effect of temperature in the range $25 \leq T \leq 230^\circ$C (typical ovens for PPLN) and report results for pump chirp $\beta=0$ and $\beta=10^{-25} ~$s$^{2}$.\\

\noindent Our results indicate that chirp effect on the spectrum depends on the sign and value, which can be observed in Fig.\ref{chirp} where $J(k_{x_s}, \omega_i)$ is shown. For comparison we included in the center the spectrum with no chirp, while to the right are collocated the results for positive and to the left for negative values of chirp. Further insight is gained from the single photon spectra, obtained by integrating $J(k_{x_s}, \omega_i)$ over $k_{x_s}$ (Eq. \ref{jpd}) and shown in the same figure. The $S(\omega_i)$ exhibits position dependence of the spectral response, as an example of the differences, the second and third row show the results for $k_{x_s}= 0$ and $k_{x_s}= 0.25 ~\mu m^{-1}$, respectively.

If instead of transverse momentum the frequency is filtered, then the crystal chirp allows to modify the emission-cone cross section. For example, removal of the central part of the energy spectrum leads to a concentric ring. This property represents a big advantage since one single crystal and the control of external parameters amounts to a source suitable for different applications.

\begin{figure}[h!]
\begin{center}
\centering\includegraphics[height=5cm, width=7.7cm]{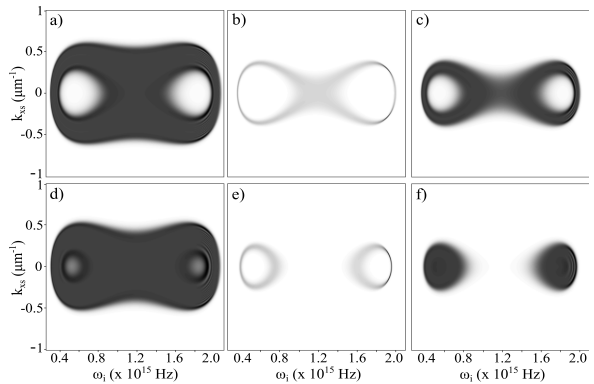}
\end{center}
\par
\caption{$J(k_{x_s}, \omega_i)$ for different temperature and chirp: The first row a ) $D=-5 \times 10^{-6} ~\mu m^{-2}$, b) $D$ = 0, and c) $D=2 \times 10^{-6} ~\mu m^{-2}$ with $T$ = 25$^{\circ}$C. Second row d), e) and f) with same values of $D$, respectively and  $T$ = 230$^{\circ}$C.} \label{temp}
\end{figure}

\noindent Temperature is a parameter that can be externally controlled and causes changes in the produced state \cite{PRL100-Nasr,JOSAB2013-Lerch}. A change in temperature induces refraction index modification and also the layer structure undergoes adjustments, for that reason temperature is commonly used to tune a sample to resonance,{\it{i.e.}} adjust it so that fulfillment of PM conditions is optimized. In Fig.\ref{temp} we report numerical results for SPDC at $T$ = 25$^\circ$C  and 230$^\circ$C and different values of crystal chirp. Notice that for zero and positive chirp values, SPDC for central frequencies (0.8 PHz $\leq \omega_i \leq$ 1.6 PHz) is not kinematically allowed, or the probability that it occurs is highly suppressed, instead the spectra portraits two separate regions with symmetric correlations  (the same result is achieved by modifying directly the central period position $z_0$ or the $\Lambda_c$ value). In contrast, for negative values of chirp, a nearly con\-ti\-nuous spectrum in frequencies and momentum is obtained.\\

\noindent Instead of presenting plots for all possible combination of variables, we will summarize our findings since there are similarities between the spectra produced when va\-ria\-tions of different parameters are considered. For exam\-ple, the effect of the crystal length is well understood for periodic poled crystals, the shorter the crystal the larger the bandwidth to the price of sacrificing the efficiency. That is one of the reasons to use chirped non-linear crystals \cite{PRA70-Carrasco, PRA81-Caspani, PRA80-Svozilik, PRA86-Svozilik}. Relevant to our discussion is the crystal chirp effect, in that case it has been shown \cite{yo2015} that chirp effects scale as $DL^2$, so that changes in the crystal length $L$ is equivalent to a change in the chirp $D$, when $D \neq 0$. Nonetheless some comments regarding $J(k_{x_s},\omega_i)$ are in order: 1) as $L$ is increased, the generated frequency and transverse momentum ranges decrease, 2) for a given $D$, for small crystal length ($L = 50 ~\mu m$), a flat spectrum is produced (see Fig.\ref{varias}a), however the pop tab-structure gradually emerges as $L$ is increased, 3) our results are in contrast with those reported in \cite{PRA81-Caspani}, where for small lengths, a structure like ours is obtained and X-entanglement for large values of L. However  neither the pop-tab structure nor the flat supercontinuum single photon spectrum are reported.

\begin{figure}[h!]
\begin{center}
\centering\includegraphics[height=5cm, width=5cm]{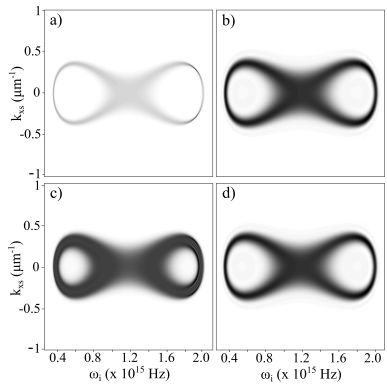}
\end{center}
\par
\caption{$J(k_{x_s}, \omega_i)$ for different chirp in pumping and in the non-linear crystal ($L$ = 5000 $\mu m$, FWHM = $0.01  ~\mu m$ and $W_{x} = W_y = 100 ~\mu m$): The first row a) $\beta=0$, b) $\beta=10^{-25} ~s^{2}$ and $D$=0. Second row, c) and d) with same $\beta$ values and $D=2 \times 10^{-6} ~\mu m^{-2}$.} \label{beta}
\end{figure}

\noindent Once the properties of the crystal are fixed, length and chirp, external parameters become relevant. By va\-rying an external parameter in crystal it is possible to obtain equivalent results to those obtained with different non-linear crystals. We already presented temperature effects, and now turn to the effect of pump properties va\-ria\-tions. We first report pump chirp effects which is readily produced propagating the pump beam through an optical fiber. The numerical results (Fig.\ref{beta}) show that some of the spectra produced in SPDC with a chirped crystal can be obtained with a PPLN crystal (without chirp) but including chirped pump (remark the strong si\-mila\-rity of the correlation in  b) and d) plots). Thus, pump chirp reveals as an effective external tool that allows to manipulate the produced state in momentum-frequency domain.

\noindent Two other parameters that can be efficiently adjusted are the spectral full width at half maximum (FWHM) and the beam waist. While the later is helpful in streng\-the\-ning the correlations in $J(k_{x_s}, k_{x_i})$, the former -larger FWHM- permits to augment the range in which the co\-rre\-la\-tion occurs. It is important to realize that in each of the layers, photons including a range of frequencies are generated, and the spectrum from different layers are distinct, even though the range of photons frequencies from neighbor layers overlap, so that when the two photon amplitude from all layers are superposed, a broad range of frequencies is covered. For small FWHM, the frequencies undergoing constructive interference is limited and the resulting pattern include areas where the intensity is weakened (see Fig.\ref{varias}c). Using this approach it is possible to obtain a variety of states ranging from highly correlated to nearly factorable states (see Fig.\ref{varias}b and c).

\begin{figure}[h!]
\begin{center}
\centering\includegraphics[height=4cm, width=7.5cm]{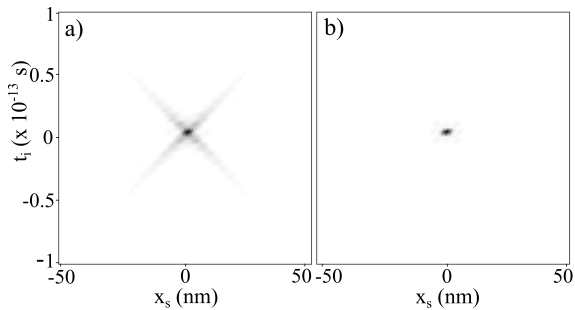}
\end{center}
\par
\caption{$\mathcal{F}J(x_s, t_i)$ for FWHM = $0.001  ~\mu m$, $L=5000 ~\mu$m, $\beta=0$, $T=25^{\circ}$C and $W_{x} = W_y = 100 ~\mu m$: a) $D$ = 0, b) $D=2 \times 10^{-6} ~\mu m^{-2}$.} \label{FT}
\end{figure}

\noindent So far we have presented our results regarding the momentum-frequency domain, we now turn to the Fourier transform of the $J(k_{x_s}, \omega_i)$, see Fig.\ref{FT}.  It is observed that, as expected, the frequency-momentum ultra-broad band leads to the strong relative localization of twin photons, both in time and position \cite{PRA86-Svozilik}. The crystal chirp effect is important to enhance the localization, for instance the $\mathcal{F}J$ Fourier transform when $D$=2 $\times 10^{-6} ~\mu m$ and FWHM=1 nm, leads to $\Delta t =6.8$ fs, and $\Delta x= 2.4$ nm, orders of magnitude  smaller (one for time, three for position)  as compared to the $D$ = 0 case. The co\-rres\-pon\-den\-ce between the temporal bandwidth and the spatial correlation length relies upon the nonseparability of the spatial and temporal domains imposed by the PM condition. Since twin photons are generated inside the non-linear medium almost simultaneously, a strong co\-rre\-la\-tion of arrival times to the crystal surface of the signal and idler photons is observed. As a consequence of dispersion properties of the non-linear medium, through which both photons at different frequencies propagate, a time-shift among detection times of both photons is produced \cite{PRA2011-Perina}.
\vspace{0.15cm}

\noindent There is a number of applications where the broadband -in the femtosecond range- and bright sources, which are produced in SPDC from chirped crystals imply a sig\-ni\-fi\-cant advantage. Some of these applications are: im\-pro\-ving resolution in quantum optical coherence tomo\-gra\-phy (QOCT) \cite{PRA65-Abouraddy, OL29-Carrasco}; designing of ultrafast correlators \cite{PRL94-Peer}; improving the accuracy of clock-synchronization and positioning measurements \cite{Nature412-Giovannetti, APL85-Valencia}; per\-for\-ming linear-optical logic operations \cite{PRA64-Grice}; improving resolution of spatial-correlation studies in colloidal systems \cite{PRL104-Peeters}; spectroscopy \cite{PRL80-Saleh}; lithography \cite{PRL85-Boto}; spatial resolved spectroscopy, like CO$_2$ detection using Laser Imaging Detection and Ranging (LIDAR) \cite{OP24-Lasse}; quantum imaging and metrology \cite{QI-Gatti}; ghost ima\-ging and diffraction experiments \cite{PRA53-Pittman}; and proving some fundamentals test in quantum mechanics \cite{MP38-Bell, NJP15-Ferrini, RMP71-Zeilinger,PRL92-Howell}. Also, by means of optical cavity filters on these sources, narrow and ultranarrow band regime is achievable, allowing their implementation in diverse quantum technologies: quantum cryptography schemes \cite{RMP74-Gisin, Nature-Duan}; optimizing biphoton-atoms intercommunication coupling \cite{PRL93-Dayan, OE15-Nielsen}; protocols like quantum teleportation \cite{PRL70-Bennet}; and distribution quantum computing \cite{PRL96-Serafini}.

\section{Summary and conclusions}

\noindent  We report the outcome from a detailed study of entangled/separables photon pairs spectra generated in  SPDC  from a LiNbO$_3$ chirped non-linear crystals when pumped at 800 nm. Numerical results are presented as plots, the typical spectrum produced in frequency-momentum space turns out to be a pop tab-like structure (see Fig.\ref{varias}). \\

\noindent {\bf{Single photon characteristics}}: Restricting our atten\-tion to one of the photons, our results -in terms of the single photon spectrum, Eq.(\ref{jpd})- show that a broadband spectra (flat supercontinuum) but also a single band or different structures are generated de\-pen\-ding upon the configuration chosen (internal (crystal chirp and length) and external  parameters (temperature, pump chirp, pump FWHM and waist)). A few examples are shown in the second and third row of Fig.\ref{chirp}.

\noindent {\bf{Twin photon characteristics}}: Pairs of photons produced in SPDC are entangled either in transverse momenta ($k_{x_s}$, $k_{x_i} $), frequency-transverse momentum  ($\omega_i$, $k_{x_s} $), time-position ($t_i, x_s$), and it is expected that high dimensional entanglement is produced in the broadband spectra. It is worth remarking the relevance of crystal chirp on the correlation between the temporal duration and the spatial localization, see Fig.\ref{FT}.

\noindent Summarizing, light source based upon SPDC from chirped crystals are well suited when broadbrand spectra or temporal and spatial resolution  are required. Furthermore, its wide multi-correlation morphology opens the possibility to fit specific-task applications.\\
\vspace{0.15cm}

\section*{Acknowledgments}

This work is supported by CONACyT under contract PDCPN2015--624 and PRODEP project UGTO-PRODEP-566.


\begin{thebibliography}{99}
\bibitem{PRA57Monken} C. H. Monken, P. H. Souto Ribeiro, S. Pad\'ua, Phys. Rev. A \textbf{57}(4), 3123 (1998).
\bibitem{Cosme} O. Cosme, A. Delgado, G. Lima, C. H. Monken, S. Pad\'ua, arXiv:0906.4734v1 (2009).
\bibitem{LP23Bobo} R. Ram\'irez-Alarc\'on, H. Cruz-Ram\'irez, A. B. U'Ren, Laser Phys. \textbf{23}, 055204 (2013).
\bibitem{yo2012} X. S\'anchez-Lozano, A. B. U'Ren and J. L. Lucio, J. Opt. 14 015202 (2012) .
\bibitem{Ansari} V. Ansari, B. Brecht, G. Harder, C. Silberhorn, arXiv:1404.7725v3 (2015).
\bibitem{OP29-Torres} Juan P. Torres, Adrian Alexandrescu, Silvia Carrasco, Lluis Torner, Opt. Lett. \textbf{29}, 4 (2004).
\bibitem{PRL101-Yu} X. Q. Yu, P. Xu, Z. D. Xie, J. F. Wang, H. Y. Leng, J. S. Zhao, Z. N. Zhu, N. B. Ming, Phys. Rev. Lett. \textbf{101}, 233601 (2008).
\bibitem{PRA86-Svozilik} J. Svozilik, J. Perina, J. P. Torres,  Phys. Rev. A \textbf{86}, 052318 (2012).
\bibitem{NJoP10-Osorio} C. I. Osorio, A. Valencia, J. P. Torres, New Journals of Physics \textbf{10}, 113012 (2008).
\bibitem{PRL109-Jedrkiewicz} O. Jedrkiewicz, A. Gatti, E. Brambilla, P. Di Trapani, Phys. Rev. Lett. \textbf{109}, 243901 (2012).
\bibitem{PRL102-Gatti} A. Gatti, E. Brambilla, L. Caspani, O. Jedrkiewicz, L. A. Lugiato, Phys. Rev. Lett. \textbf{102}, 223601 (2009).
\bibitem{PRA82-Brambilla} E. Brambilla, L. Caspani, L. A. Lugiato, A. Gatti, Phys. Rev. A \textbf{82}, 013835 (2010).
\bibitem{PR85-Brambilla} E. Brambilla, O. Jedrkiewicz, L. A . Lugiato, A. Gatti, Phys. Rev. A \textbf{85}, 063834 (2012).
\bibitem{IJQI12-Gatti} A. Gatti, L. Caspani, T. Corti, E. Brambilla, O. Jedrkiewicz, Int. J. Quantum Inf. \textbf{12} (2), 1550032 (2014).
\bibitem{PRA81-Caspani} L. Caspani, E. Brambilla, A. Gatti, Phys. Rev. A \textbf{81}, 033808 (2010).
\bibitem{PRA73-Carrasco} Silvia Carasco, Alexander V. Sergienko, Bahha E. A. Saleh, Malvin C. Teich, Phys. Rev. A \textbf{73}, 063802 (2006).
\bibitem{yo2015} X. S\'anchez-Lozano, J. L. Lucio M., Int. J. Quantum Inf. \textbf{13} (5), 1550032 (2015).
\bibitem{PRA80-Svozilik} J. Svozil\'ik, J. Perina, Phys. Rev. A \textbf{80}, 023819 (2009).
\bibitem{PRL98-Harris} S. E. Harris, Phys. Rev. Lett. \textbf{98}, 063602 (2007).
\bibitem{PRL100-Nasr} M. B. Nasr, S. Carrasco, B. E. Saleh, A. V. Sergienko, M. C. Teich, J. P. Torres, L. Torner, D. S. Hum, M. M. Fejer, Phys. Rev. Lett. \textbf{100}, 183601 (2008).
\bibitem{PRA2011-Perina} J. Perina, J. Svozilik, Phys. Rev. A \textbf{83}, 033808 (2011).
\bibitem{JOSAB2013-Lerch} S. Lerch, B. Bessire, c. Bernhard, T. Feurer and A. Stefanov, J. Opt. Soc. Am. B \textbf{30} (4), 953 (2013).
\bibitem{PRA70-Carrasco} Silvia Carrasco, Juan P. Torres and Lluis Torner, Phys. Rev. A \textbf{70}, 043817 (2004).
\bibitem{PRA65-Abouraddy} A. F. Abouraddy, M. B. Nasr, B. E. A. Saleh, A. V. Sergienko, M. C. Teich, Phys. Rev. A \textbf{65}, 053817 (2002).
\bibitem{OL29-Carrasco} S. Carrasco, J. P. Torres, L. Torner, A. Sergienko, B. Saleh, M. Teich, Opt. Lett. \textbf{29}, 2429 (2004).
\bibitem{PRL94-Peer} A. Pe\'er, B. Dayan, A. A. Friesem, Y. Silberberg, Phys. Rev. Lett. \textbf{94}, 073601 (2005).
\bibitem{Nature412-Giovannetti} V. Giovannetti, S. Lloyd, L. Maccone, Nature \textbf{412}, 417 (2001).
\bibitem{APL85-Valencia} A. Valencia, G. Scarcelli, Y. H. Shih, Appl. Phys. Lett. \textbf{85}, 2655 (2004).
\bibitem{PRA64-Grice} W. P. Grice, A. B. URen, I. A. Walmsley, Phys. Rev. A \textbf{64}, 063815 (2001).
\bibitem{PRL104-Peeters} W. H. Peeters, J. J. D. Moerman, M. P. van Exter, Phys. Rev. Lett. \textbf{104}, 173601 (2010).
\bibitem{PRL80-Saleh} B. E. A. Saleh et al., Phys. Rev. Lett. \textbf{80}, 3483 (1998).
\bibitem{PRL85-Boto} A. N. Boto et al., Phys. Rev. Lett. \textbf{85}, 2733 (2000).
\bibitem{OP24-Lasse} Lasse H$\o$gstedt, Andreas Fix, Martin Wirth, Christian Pedersen, Peter Tidemand-Lichtenberg, Optics Express \textbf{24}, 5, 254503 (2016).
\bibitem{QI-Gatti} A. Gatti, E. Brambilla, Lugiato, Quantum Imaging (Elsevier,North-Holland, 2008).
\bibitem{PRA53-Pittman} T. B. Pittman, D. V. Strekalov, D. N. Klyshko, M. H. Rubin, A. V. Sergienko, Y. H. Shih, Phys. Rev. A \textbf{53}, 2804 (1996).
\bibitem{MP38-Bell} J. S. Bell, Rev. Mod. Phys. \textbf{38}, 447-452 (1966).
\bibitem{NJP15-Ferrini} G. Ferrini, J. P. Gazeau, T. Coudreau, C. Fabre, N. Treps, New Journal of Physics \textbf{15}, 093015 (2013).
\bibitem{RMP71-Zeilinger} A. Zeilinger, Rev. Mod. Phys. \textbf{71}, S288 (1999).
\bibitem{PRL92-Howell} J. C. Howell, R. S. Bennink, S. J. Bentley, R. W. Boyd, Phys. Rev. Lett. \textbf{92}, 210403 (2004).
\bibitem{RMP74-Gisin} N. Gisin, Rev. Mod. Phys. \textbf{74}, 145 (2002).
\bibitem{Nature-Duan} L. M. Duan, M. D. Lukin, J. I. Cirac, P. Zoller, Nature \textbf{414}, 413 (2001).
\bibitem{PRL93-Dayan} B. Dayan, A. Pe\'er, A. A. Friesem, Y. Silberberg, Phys. Rev. Lett. \textbf{93}, 023005 (2004).
\bibitem{OE15-Nielsen} J. S. Neergaard-Nielsen et al., Opt. Express \textbf{15}, 7940 (2007).
\bibitem{PRL70-Bennet} C. H. Bennett, G. Brassard, C. Cr\'epeau, R. Jozsa, A. Peres, W. K. Wootters, Phys. Rev. Lett. \textbf{70}, 1895 (1993).
\bibitem{PRL96-Serafini} A. Serafini, S.Mancini, S. Bose, Phys. Rev. Lett. \textbf{96}, 010503 (2006).
\end{thebibliography}
\end{document}